\documentclass[aps,prb,superscriptaddress,draft,showpacs,intlimits,amsmath,amssymb,floats,floatfix,twocolumn]{revtex4}
\usepackage{bm}
\usepackage[final]{graphicx}
\usepackage[thinspace]{SIunits}
\usepackage{color}

\newcommand{\wn}{\mbox{$\,\mathrm{cm^{-1}}$ }}               % wavenumber
\newcommand{\NZ}{\mbox{$\mathrm{Ni_{67}Zr_{33}}$ }}                % Ni67Zr33

\begin{document}
\title{Light scattering study of low-energy vibrational excitations in the metallic glass Ni$_{67}$Zr$_{33}$ using electronic Raman scattering}
\date{\today}
\author{B. Muschler}
\affiliation{Walther Meissner Institut, Bayerische Akademie der Wissenschaften, 85748 Garching, Germany}
\author{I. T\"utt\H{o}}
\affiliation{{Research Institute for Solid State Physics and Optics, Hungarian Academy of Sciences, 1525 Budapest, Hungary}}
\author{A. Zawadowski}
\affiliation{Institute of Physics, Budapest University of Technology and Economics, 1111 Budapest, Hungary}
%\author{G. Mih\'aly}
%\affiliation{Institute of Physics, Budapest University of Technology and Economics, 1111 Budapest, Hungary}
\author{J. Balogh}
\affiliation{{Research Institute for Solid State Physics and Optics, Hungarian Academy of Sciences, 1525 Budapest, Hungary}}
\author{R. Hackl}
\affiliation{Walther Meissner Institut, Bayerische Akademie der Wissenschaften, 85748 Garching, Germany}
%\email{hackl@wmi.badw.de}

%%%%%%%%%%%%%%%%%%%%%%%%%%%%%%%%%%%%%%%%%%%%%%%%%%%%%%%%%%%%%%%%%%%%%%%%%%%%%%%%%%%%%%%%%
\begin{abstract}
The Raman response of the metallic glass \NZ is measured as a function of polarization and temperature and analyzed theoretically. Unexpectedly, the intensity in the range up to 300\wn increases upon cooling, which is counterintuitive when the response originates from vibrations alone as in insulators. The increase finds a natural explanation if the conduction electrons are assumed to scatter on localized vibrations with a scattering probability proportional to the Debye-Waller factor. None of our assumptions is material specific, and the results are expected to be relevant for disordered systems in general.
\end{abstract}
%%%%%%%%%%%%%%%%%%%%%%%%%%%%%%%%%%%%%%%%%%%%%%%%%%%%%%%%%%%%%%%%%%%%%%%%%%%%%%%%%%%%%%%%%
\maketitle
%%%%%%%%%%%%%%%%%%%%%%%%%%%%%%%%%%%%%%%%%%%%%%%%%%%%%%%%%%%%%%%%%%%%%%%%%%%%%%%%%%%%%%%%%
\section{Introduction}

The intriguing properties of disordered or amorphous materials have been studied in great detail \cite{Glassy:1981} and play an important role in many contemporary studies and applications. From the viewpoint of microscopic properties, one of the hallmarks is an excess density of states (DOS) in the vibrational spectrum as well as extra contributions to the low-temperature heat capacity and conduction and to the ultrasound attenuation \cite{Glassy:1981}. The extra vibrational DOS which is superposed on the Debye-like $ \omega^{2} $ spectrum of crystals is usually referred to as the boson peak. In amorphous insulators a random distribution of the elastic constants is appropriate to capture the experimental results \cite{Schmid:2008}. However, local modes, believed to be relevant for both disordered crystals and amorphous materials, can be the origin of the boson peak \cite{Maurer:2004}. In metallic Ni-B, which exists in crystalline and amorphous form,  extra vibrational states were observed in a neutron scattering study at both low and high energies \cite{Lustig:1986}. In fact, the vibrations in disordered systems can be phonons with considerable energies as well as low-lying excitations in the meV range usually present when one or a few atoms have more space and vibrate in voids, described as a potential well. The latter excitations can be quite localized.  The two lowest levels of these local modes are frequently called two level systems (TLS) and play an important role in the low-temperature thermodynamic behavior.

Raman scattering experiments were instrumental in studying the vibrational properties of amorphous materials since the complete spectrum can be observed because of the absence of translational invariance and the related collapse of the selection rules. In an early calculation the corresponding Raman response $\chi^{\prime\prime}_{i,s}(\omega,T)$ was found to be proportional to the vibrational DOS $ g(\omega) $ as \cite{Shuker:1970}
\begin{equation}
  \frac{S_{i,s}(\omega,T)}{1+n(\omega,T)} = \chi^{\prime\prime}_{i,s}(\omega,T) = c_{i,s}\frac{g(\omega)}{\omega}
  \label{eq:SG}
\end{equation}
where $n(\omega,T)= [e^{\hbar\omega/k_BT}-1]^{-1}$ is the Bose-Einstein occupation number at temperature $ T $, $ S_{i,s} (\omega)$ is the measured Raman intensity (structure factor), and $\omega=\omega_{i} -\omega_{s} $ is the energy transferred to the system by incoming and scattered polarized photons $i$ and $s$, respectively. The coupling factors $c_{i,s}$, originally assumed to be constant, turned out to have a rather complicated energy dependence \cite{Jackle:1981,Surovtsev:2002}. The factor $\omega^{-1}$ is separated out of the coupling factor as the most probable energy dependence due to the matrix elements of the harmonic oscillators. Schmid and Schirmacher showed recently for insulators that $ g(\omega) $ is not simply the DOS. Rather, $\chi^{\prime\prime}_{i,s}(\omega,T)$ is a polarization dependent weighted sum of the longitudinal and the transversal susceptibilities in systems with fluctuating elasto-optical coupling and force constants.\cite{Schmid:2008} A similarly rigorous treatment for metals is still missing.

In this paper we analyze Raman scattering results of the metallic glass \NZ and show that there are three relevant types of excitations which can be disentangled. There is a temperature-independent Drude-like response due to the electron-hole excitations as suggested earlier for dirty metals \cite{Zawadowski:1990}. On top of the electronic spectrum, there is an additional response which, at 300\,K, is strictly linear up to 20\wn. In the range 20 to 300\wn, there is a wide maximum which loses about 30\% of its spectral weight upon increasing the temperature, just opposite what is expected from the occupation number of bosonic excitations. To our knowledge only the Debye-Waller factor may account for such behavior by reducing the scattering probability of electrons on vibrations. Therefore, substantial vibrational density of states is required at relatively low energies to produce a strong enough temperature dependence. We focus on this contribution.

\section{Samples and Experiment}

The amorphous \NZ sample was prepared by melt-spinning 99.9\% pure Zr
and Ni metals in an Ar atmosphere after electron-beam melting
in a quartz tube. The non-crystalline state was checked by x-ray
diffraction. The amorphous character is also reflected in the slight (1\%)
increase of the resistivity $\rho(T)$ upon cooling from room temperature to 4\,K, which is just opposite what is observed in microcrystalline samples \cite{Altounian:1983}. A mean free path $\ell$ of approximately 20\,\AA ~or some $5--10$ interatomic distances $a$ is estimated from the resistivity, $\rho = 150\,\mu\Omega$\,cm, and the density of states at the Fermi level $N(E_F)$ \cite{Bakonyi:1995}.

The Raman experiments were performed with standard equipment. For excitation a solid state laser emitting at 532\,nm (Klastech Scherzo-Denicafc-532-300) was used. The spectra are corrected for the sensitivity of the instrument. We always show the response function $R\chi^{\prime\prime}_{i,s}(\omega,T) \propto S_{i,s}/\{1+n(\omega,T)\}$. $R$ absorbs all experimental factors and  $\chi^{\prime\prime}(\omega,T)_{i,s}$ is the imaginary part of the dimensionless Raman response function.

\section{Results and Discussion}

\begin{figure}
  \centering
  \includegraphics[width=8cm]{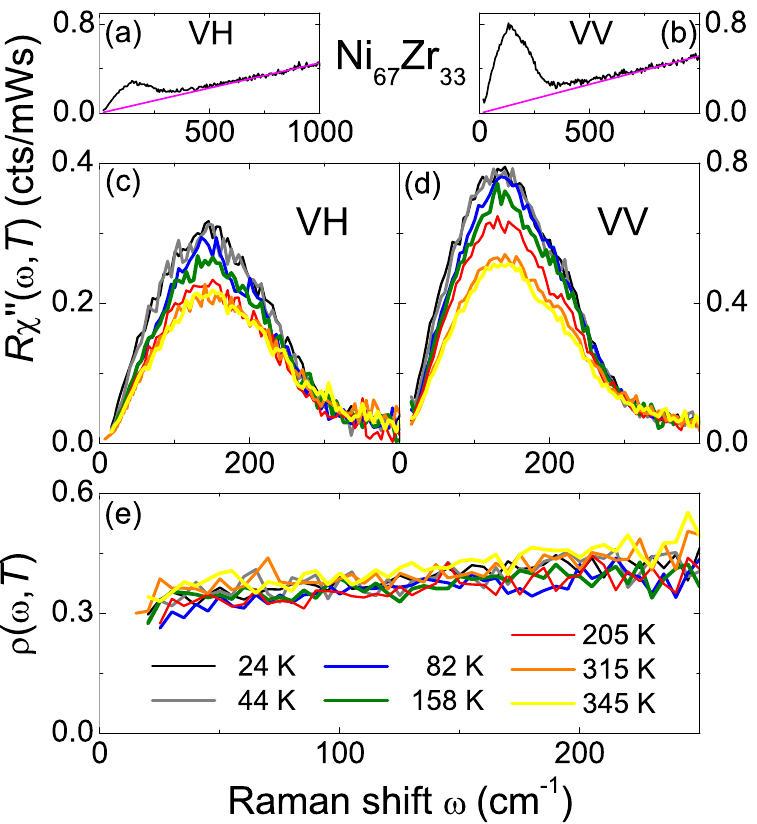}
  %\vspace{-0.8cm}
  \caption[]{(Color online)  Raman response $R\chi^{\prime\prime}(\omega,T)$ of \NZ for perpendicular (VH) and parallel (VV) light polarizations. (a) and (b) show on an extended energy range that the spectra consist of a polarization independent linear continuum and superposed peaks in the range below 300\wn. (c), (d) After subtraction of the linear part, the low-energy peaks have nearly identical shapes for all temperatures [c.f. (e)], but the overall intensity in VH is only approximately 30\% of that in VV. (e) The ratio $\rho=\chi^{\prime\prime}_{VH}/\chi^{\prime\prime}_{VV}$ is temperature independent and increases slightly with energy.
  } \label{fig:raw}
\end{figure}

In Fig.~\ref{fig:raw}, the temperature dependence of the Raman spectra of \NZ with parallel (VV) and perpendicular (VH) light polarizations is shown. There is a slightly asymmetric maximum centered at about 140\wn (1\,meV\,=\,8\wn) which is superposed on a linear continuum extending to several 100\,meV [Figs.~\ref{fig:raw}~(a) and \ref{fig:raw}(b)]. While the continuum is temperature independent the intensity in the peak maximum increases by approximately 30\% upon cooling [Figs.~\ref{fig:raw}~(c) and \ref{fig:raw}(d)].

%%%%%%%%%%%%%%%%%%%%%%%%%%%%%%%%%--Fig--1--%%%%%%%%%%%%%%%%%%%%%%%%%%%%%%%%%%%%%%%%%%%%%%%%%%%%%%%%
%%%%%%%%%%%%%%%%%%%%%%%%%%%%%%%%%%%%%%%%%%%%%%%%%%%%%%%%%%%%%%%%%%%%%%%%%%%%%%%%%%%%%%%%%

The continuum has a strong resemblance to the behavior of dirty metals for which the theory was worked out years ago \cite{Zawadowski:1990}. In that work different scattering channels classified by crystal harmonics $ L $  were considered. The electron density fluctuations that appear in the channel $ L=0 $ are screened and cannot contribute to the spectrum. However, the channels $ L=2n > 0 $ are relevant implying that the electrons are being redistributed between different parts of the Fermi surface. Only those particles contribute to the cross section which are scattered out of a given channel. The scattering mechanism is similar to that in multi-valley semiconductors \cite{Ipatova:1981}. In \NZ the continuum can be explained essentially in this way with a relaxation rate in the range of 0.5\,eV, consistent with the resistivity of $150\,\mu\Omega$\,cm. The position of the low-energy peak would correspond to a much longer relaxation time (smaller rate). In addition, the spectral shape deviates from the expected one \cite{Zawadowski:1990}. Hence, the peak has to be interpreted differently.

Following the reasoning for insulators, one can argue that the peak resembles that of vibrational properties \cite{Shuker:1970,Schmid:2008}. Schmid and Schirmacher \cite{Schmid:2008} make a prediction for the ratio of the responses in the VH and VV configurations, $\rho(\omega,T)=\chi^{\prime\prime}_{VH}/\chi^{\prime\prime}_{VV}$, derived for the case of locally varying elastic constants. In our experiment, $\rho(\omega,T)$ increases with energy as shown in Fig.~\ref{fig:raw}~(e) rather than decreases as predicted. Whether or not the discrepancy can be traced back to the metallicity of \NZ cannot be answered in the present state of knowledge. In addition, in the case of a vibrational origin the intensity of the peak should increase rather than decrease with $ T $ since the occupation number of vibrations with energy $\omega$ is proportional to $n(\omega,T)$. In crystals with well defined electronic levels having widths in the range of $k_BT$, resonance effects could reverse the temperature dependence expected for vibrations by overcompensating the effect of the Bose factor \cite{Panfilov:1998,Tassini:2008}. However, in an amorphous material with electronic relaxation rates in the range of eV, as directly derived from the Raman experiment (see above), the resonances are too wide to be relevant. In fact, the electronic part is independent of the temperature. We have also verified experimentally that both the high- and low-energy contributions are independent of the energy of the exciting photons. Finally, the electrons could scatter from spin fluctuations which gain influence toward lower temperature in a paramagnetic solid. However, for Ni concentration below 70\,\% the paramagnetism is very weak \cite{Bakonyi:1995} as corroborated by our own magnetization measurements for the sample used here. Moreover, Batalla {\it et al.} \cite{Batalla:1985} concluded from their studies of the low-temperature specific heat that the coupling to spin fluctuations in Ni-Zr compounds is negligible in the composition range of interest here. Since, to the best of our knowledge, studies of the specific heat or of the vibrational spectra by other methods do not exist in the temperature range covered here further insight into the interpretation of the Raman results from a comparison to other work is not possible at present.

\begin{figure}[tb]
  \centering
  \includegraphics[width=7cm]{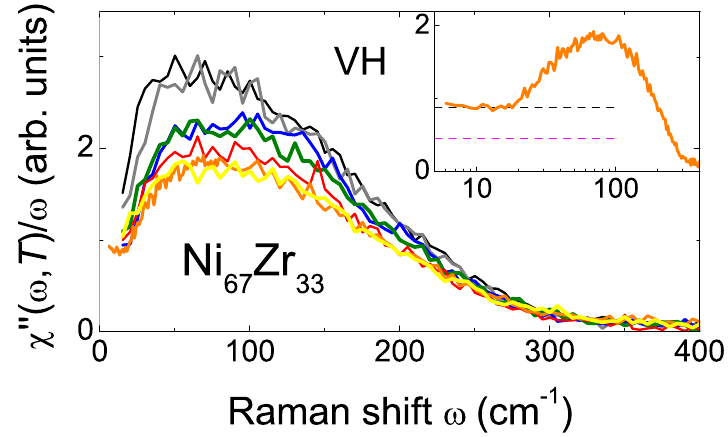}
  \caption{(Color online) ``Boson'' peak in \NZ [same temperatures as in Fig.~\ref{fig:raw}~(c-e)]. The inset shows the spectrum at 315\,K, which was measured down to 6\wn, on  a logarithmic scale. The slight increase toward $\omega \rightarrow 0$ is a remainder of the laser line. At the other temperatures the spectra were measured only for $\omega \ge 15$\wn. In the plot $\chi^{\prime\prime}/\omega$ a Drude-like spectrum appears as a constant (lower dashed line in the inset). The measured low-energy spectrum saturates above the Drude response (upper dashed line), indicating scattering which could originate from vibrations with a Debye-like spectrum (see text).}
  \label{fig:reduced}
\end{figure}

It is instructive to divide the low-energy response [Figs.~\ref{fig:raw}~(c) and \ref{fig:raw}(d)] by $\omega$ as shown for the VH case in Fig.~\ref{fig:reduced}. In the VH spectrum at 315\,K, which was measured with high resolution, a constant part on top of the Drude response emerges below 20\wn  as highlighted in the inset. In the spirit of Eq.~(\ref{eq:SG}) the constant part would correspond to a Debye-like $\omega^2$ variation of the DOS $g(\omega)$ if the coupling factors $c_{i,s}$ were constant. However, in a large number of studies on insulators and semiconductors, different energy dependences for $c_{i,s}$ are found \cite{Jackle:1981,Surovtsev:2002}. Further clarification of this question must be postponed to another publication. Above 20\wn, additional scattering sets in abruptly which, for the various reasons just discussed, cannot be interpreted in the same way as the boson peak in insulators. The increase toward low $ T $ indicates that other types of excitations must contribute to the inelastic relaxation of the electrons.

In the following we study the electron relaxation on static defects such as impurities and on vibrations at energies in the range of and below a maximal vibration frequency $ \omega \simeq \omega_{0}$ on the order of the Debye energy. The thermal occupation number can affect the Drude peak and lead to an increase of the scattering amplitude in the spectrum with temperature. Given the observed temperature dependence this process cannot be the dominating one. In the range $20--300$\wn, an additional peak shows up on top of the Drude background. Electron-hole pairs are created which are coupled to local vibrations. If low-lying vibrations at $ \omega\ll \omega_{0}$ are created the amplitude $u$ of the vibrations can be large. We assume that $ u $ can be approximated by the result for the harmonic oscillator, $|u(\omega)|^{2}=(\hbar/M\omega)\lbrace1/2+ n(\omega,T)\rbrace$, where $ M $ is the relevant mass. Since the electron has a momentum close to the Fermi momentum, $k_{F} \simeq \pi/a$, the situation is comparable to that of x-ray diffraction on crystals, and the scattering probability increases at lower temperatures according to the Debye-Waller factor, $e^{-2W(\omega,T)}$, with
$2W(\omega,T)=\vert \bf{k} \cdot \bf{u} \vert^{2}$. In other words, the local oscillator is a weaker scatterer at high
temperature when it is blurred out because of its large amplitude and a strong one at low temperatures when it is almost point-like. Since $n(\omega,T)$
%%%%%%%%%%%%%%%%%%%%%%%%%%%%%%%%%%%%%%%%%%%%%%%%%%%%%%%%%%%%%%%%%%%%%%%%%%%%%%%%%%%%%%%%%
%%%%%%%%%%%%%%%%%%%%%%%%%%%%%%%%%%%%%%%%%%%%%%%%%%%%%%%%%%%%%%%%%%%%%%%%%%%%%%%%%%%%%%%%%
also appears in the exponent, the Debye-Waller factor could dominate over the occupation number at low energies as observed in the experiments (Fig.~\ref{fig:raw}). On the other hand, static impurities and other imperfections of the system may be very rigid; thus their Debye-Waller factor is small. This is in accordance with the temperature--independent linear continuum.

The Debye-Waller factor for a single vibration with energy $ \omega_{j} $ can be estimated as $2W_{j}=(\alpha/\omega_{j})\left\{ 1/2 + n(\omega_{j},T)\right\}$, with $ \alpha=2k^{2}(\hbar/M) $. $ M $ is the mass of the vibrating atom, and $k \simeq k_{F} \simeq \lambda(\pi/a)$ where $\lambda$ is a number of order unity. Thus, $\alpha$ can be estimated as $\alpha \simeq 4\lambda^2\omega_F(m/M)$ ($\omega_F$ has the same units as $\omega_j$). With the magnitude of the Fermi energy $E_F = \hbar\omega_F \simeq 5$\,eV, $\alpha$ will be on the order $2\lambda^2$\,(cm$^{-1}$).

Now, the creation of a vibration at $\omega_{j}$ by the electron-hole continuum in the presence of static impurities and disorder is considered \cite{Zawadowski:1990}. The electron-hole continuum arising in channel $ L $ is described by the density operator $\tilde{\rho}_{L}(q)=\sum_{k}\gamma_{L} (k)a_{k}^{+}a_{k+q}$ and the corresponding correlation function is $\pi^{0}_{L}(\omega)=-2N_{F} \vert \gamma_{L}\vert^{2}\left( 1-i\omega\tau^{\ast}_{L}\right)^{-1}$, where $ N_{F}$ is the density of states at the Fermi level, $ \gamma_{L} $ is the amplitude of the photon-electron-hole coupling and $ {\tau^{\ast}_{L}}^{-1} $ is the relaxation rate. In the following we approximate $ \tau^{\ast}_{L} $ with $ \tau_{0}$ as they are not very different. The Hamiltonian of the vibration is $ \sum_{j}\omega_{j}b_{j}^{+}b_{j} $, where $  \omega_{j}$, $ b_{j}^{+} $, and $ b_{j}$ are the energy and the creation and annihilation operators of the phonon, respectively. The vibrational Green's function is given by the usual propagator multiplied by an exponential term due to the temperature dependent scattering probability,
\begin{equation}
    D_j(\omega) = \frac{\omega_j}{\omega^2-\omega_j^2}e^{-2W(\omega_j,T)}
    \label{eq:loc}
\end{equation}
The electron continuum is coupled to the vibrations by a local interaction, similar to that in small-polaron theory, as $ \sum_{j} g_{L,j}\tilde{\rho}_{L}(R_{i})\left(b_{j}^{+}+b_{j}\right)$, where $ g_{L,j} $ is the coupling and $\tilde{\rho}_{L}(R_{i}) $ is the electron density operator at the position $R_{i} $ of the local oscillator. The actual coupling is very model dependent but in order to demonstrate the anomalous temperature dependence we restrict ourselves to the simple Hamiltonian above. After averaging over the positions of the static impurities $\pi^{0}_{L} $ becomes independent of the position, and the renormalized correlation function is $ \pi_{L}(\omega)=\pi^{0}_{L}(\omega)\left[1-\sum_{j} g_{L,j}^{2}D_{j}(\omega)\pi^{0}_{L}(\omega)\right]^{-1}$.

\begin{figure}[tb]
  \centering
  \includegraphics[width=7cm]{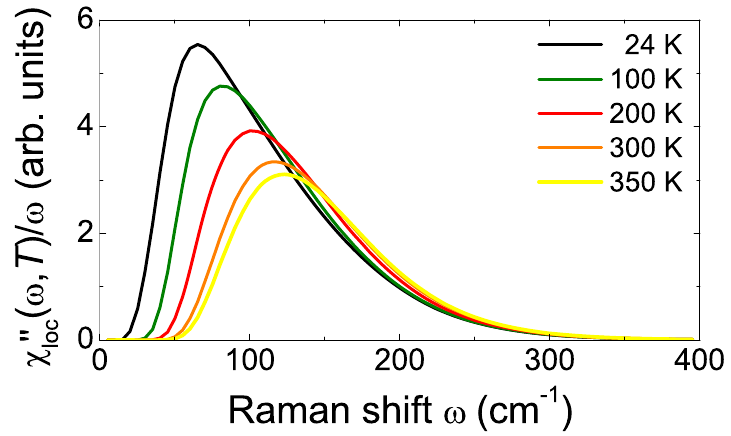}
  \caption{(Color online) Response from localized oscillators $\chi^{\prime\prime}_{\rm loc}/\omega$. The low-energy cut-off due to the Debye-Waller factor is moving downward with decreasing temperature. At room temperature the cut-off can be clearly resolved in the experiments (inset of Fig.~\ref{fig:reduced}).}
  \label{fig:loc}
\end{figure}

The strength of the dimensionless coupling can be estimated as $ N_{F}\sum_{j} g_{L,j}^{2}D_{j}(\omega)\sim N_{F}N|V|^{2}/\tilde{\omega}_{j}$, where $  N$ is the number of the active vibrational centers per volume, $ \tilde{\omega}_{j} $ is their typical energy ($ 10 \leq \tilde{\omega_{j}}\leq 100\,{\rm K} $) and $ V $ is the coupling potential. For a rough estimate typical values are borrowed from the corresponding estimates for TLSs \cite{Vladar:1983}: $ |VN_{F}| \leq 0.2$ and $ N\approx N(\epsilon)\tilde{\omega}_{j} $, where $ N(\epsilon) $ is the energy density of the centers, e.g., $ N(\epsilon)\approx 5 \times 10^{-6}\,K^{-1}\,{\rm cm}^{-3}= 5\,{\rm ppm\,K}^{-1}$. Thus the order of magnitude at the relevant energy window of that dimensionless coupling is $ N_{F}N|V|^{2}/\tilde{\omega}_{j} \sim N(\epsilon)|VN_{F}|^{2}/N_{F} \leq 5\times 10^{-2}$. Assuming weak coupling the summation over the perturbation series is not required. Keeping first order only, $\pi_{L}(\omega)\approx\pi^{0}_{L}(\omega)\left[1+\sum_{j} g_{L,j}^{2}D_{j}(\omega)\pi^{0}_{L}(\omega)\right]$, and the Raman response for the localized modes alone reads
\begin{eqnarray}
   \chi^{\prime\prime}_{\rm loc}(\omega,T)&=& {2\pi (N_F)^2|\gamma_L|^2}  \nonumber\\
   &\times& g_{\rm loc}(\omega)\{ 1+n(\omega,T)\}e^{-2W(\omega,T)}
   \label{eq:response}
\end{eqnarray}
and is plotted in Fig. \ref{fig:loc}.
Equation (\ref{eq:response}) is valid in the limit $ \omega\tau_0 \ll 1 $ and for $ \omega\geq 0$, where vibrations are created, and $g_{\rm loc}(\omega)\{ 1+n(\omega,T)\}e^{-2W(\omega,T)}=-\pi^{-1} Im\left\{\sum_{j}g_{L,j}^{2}D_{j}(\omega)\right\}$. Here, $g_{\rm loc}(\omega)$ is proportional to the effective density
of states of the local oscillators, which is weighted by the occupation number and the Debye-Waller factor. For $  \omega > \omega_0$, $ g_{\rm loc}(\omega) $ falls off, and at larger energies only the Drude peak survives. In the temperature dependence of the low-lying vibrational contribution, the Debye-Waller factor dominates over the temperature dependence of the occupation number $ n(\omega,T) $, and the low-energy part is suppressed as it is observed experimentally. The residual scattering below 20\wn on top of the Drude response [Fig.~\ref{fig:reduced} inset] will be studied in more detail in a future presentation.

For the Debye-Waller factor to become effective, a substantial density of states at energies typically below 100\wn is required. The usual Debye-like spectrum  with the DOS starting out as $\omega^2$ is inappropriate. TLSs, on the other hand, have excitations with extremely small energies. Since the temperatures of interest are relatively high we can mimic the potential as a single well with a harmonic spectrum and an occupation $n  \gg 1$. In order to produce spectra we assume a constant $ g_{\rm loc}(\omega)$ as the simplest example. We emphasize that there may be a substantial energy dependence, the discussion of which, however, is beyond the scope of this publication. Here, $ g_{\rm loc}(\omega)$ is cut off smoothly at low and high energies by hyperbolic tangent functions. As the low-energy vibrations are suppressed by the Debye-Waller factor, the results are insensitive to the low-energy cut off. In addition to the TLSs, other types of excitation may exist at very low energies. The actual calculation follows these steps: $\tau_0$ is  fitted to describe the Drude peak at higher energies, and $\omega_{0}$ is determined by the width of the vibrational peak. The effective density of localized states $g_{\rm loc}$ is constant with lower- and higher-energy soft cut offs at 40 and 160\wn with 20 and 90\wn widths, respectively, and $\int{d\omega g_{\rm loc}}=1$. Also, $\alpha = 80$\wn is relatively large. However, the localized modes vibrate in voids with a very flat potential at equilibrium, and the harmonic approximation underestimates the occupation number substantially. In addition, the momentum transfer $k$ may be larger than $\pi/a$, and the parameter $\lambda$ can easily be between 2 and 4.

%%%%%%%%%%%%%%%%%%%%%%%%%%%%%%%%%%%%%%%%%%%%%%%%%%%%%%%%%%%%%%%%%%%%%%%%%%%%%%%%%%%%%%%%%
%%%%%%%%%%%%%%%%%%%%%%%%%%%%%%%%%%%%%%%%%%%%%%%%%%%%%%%%%%%%%%%%%%%%%%%%%%%%%%%%%%%%%%%%%

Finally, the relative insensitivity of the spectral shapes at low energy to the light polarizations [see Figs.~\ref{fig:raw}~(c)--\ref{fig:raw}(e)] requires a comment since there is a discrepancy with the results of insulators \cite{Schmid:2008}. In amorphous metals, due to the relatively short mean free path of the electrons, the intensities depend only on the light-electron bare vertex functions that contain matrix elements of the form $\langle F|\hat{\bf p}{\bf e}_s|\nu\rangle\langle \nu|\hat{\bf p}{\bf e}_i|I\rangle$ with $I,\,F,\,\nu$ the initial, final and intermediate states, respectively, ${\hat{\bf p}}$ the momentum operator, and ${\bf e}_{i,s}$ the photon polarizations.  In the tight binding approximation the intra-atomic matrix elements are the most important ones and can be transformed into dipole matrix elements. Hence, the selection rules for intra-atomic orbitals apply and can explain the weaker intensity of the VH component (crossed polarizations).\cite{Pockels:2011} For completeness, we should mention that there is an additional peak at a few eV which is attributed to the multiple electron-hole excitations \cite{future:2011}.

\section{Conclusions}

In conclusion, we have shown experimentally and theoretically that the low-energy part of the Raman spectra in the metallic glass \NZ originates from the superposition of electronic relaxation processes due to scattering from impurities and from localized vibrations. The temperature dependence of the spectral weight indicated that local modes contribute substantially to the cross section. The increase with decreasing temperature can in fact be described in a qualitative way assuming that the scattering of conduction electrons is dominated by the temperature dependence of the Debye-Waller factor rather than by the occupation number of the vibrations alone. For a quantitative comparison of experiment and theory, a concrete assumption for the effective density of states would be needed. As pointed out earlier, this would be very model dependent but would not influence the main conclusion concerning the influence of the Debye-Waller factor. The result may be applicable to a large variety of systems such as doped semiconductors, metallic functionalized carbon nanotubes, and polymers.

\section{Acknowledgments}

We gratefully acknowledge discussions with W. Schirmacher and G. Mih\'aly.
We also thank A. Lovas for providing the sample, A. Geresdi for the
resistivity data, and M. Opel for the magnetization measurements. The work was supported by the DFG [Trans\-regional Research Center TRR\,80 and Research Unit FOR\,538 (Grant-No. HA2071/3)], OTKA (Grant-No. CNK\,80991), OMFB (Grant-No. 00588/2010), and the Alexander von Humboldt Foundation (Grant-No. 3-FoKoop-DEU/1009755).

%\bibliography{E:/!papers/!bib/literatureR}

\begin{thebibliography}{10}
\bibitem{Pockels:2011}
    There is an additional argument dealing with the difference between metals and insulators. In an insulator the light couples to the lattice via the valence electrons that determine the the force constants. In a metal, the light couples predominantly to the conduction electrons, and at least part of the spectral weight of vibrations results from a finite electron-phonon coupling $\lambda$. The other part of the spectral weight of phonons originates from the excitation via valence electrons as in insulators. The total weight and the ratio depend on the symmetry of the phonons and on the screening.
\end{thebibliography}
%\bibliographystyle{prsty}

%\end{document}

\end{document}